%
\let\includefigures=\iftrue
%
%
\batchmode
  \font\blackboard=msbm10
\errorstopmode
\newif\ifamsf\amsftrue
\ifx\blackboard\nullfont
  \amsffalse
\fi
\newfam\black
%
%
%
\input harvmac.tex
\includefigures
\message{If you do not have epsf.tex (to include figures),}
\message{change the option at the top of the tex file.}
\input epsf
\def\figin{\epsfcheck\figin}\def\figins{\epsfcheck\figins}
\def\epsfcheck{\ifx\epsfbox\UnDeFiNeD
\message{(NO epsf.tex, FIGURES WILL BE IGNORED)}
\gdef\figin##1{\vskip2in}\gdef\figins##1{\hskip.5in}
\else\message{(FIGURES WILL BE INCLUDED)}%
\gdef\figin##1{##1}\gdef\figins##1{##1}\fi}
\def\DefWarn#1{}
\def\figinsert{\goodbreak\midinsert}
\def\ifig#1#2#3{\DefWarn#1\xdef#1{fig.~\the\figno}
\writedef{#1\leftbracket fig.\noexpand~\the\figno}%
\figinsert\figin{\centerline{#3}}\medskip\centerline{\vbox{\baselineskip12pt
\advance\hsize by -1truein\noindent\footnotefont{\bf Fig.~\the\figno:} #2}}
\bigskip\endinsert\global\advance\figno by1}
\else
\def\ifig#1#2#3{\xdef#1{fig.~\the\figno}
\writedef{#1\leftbracket fig.\noexpand~\the\figno}%
\global\advance\figno by1}
\fi
\ifamsf
\font\blackboards=msbm7
\font\blackboardss=msbm5
\textfont\black=\blackboard
\scriptfont\black=\blackboards
\scriptscriptfont\black=\blackboardss
\def\Bbb#1{{\fam\black\relax#1}}
\else
\def\Bbb{\bf}
\fi
%
\def\yboxit#1#2{\vbox{\hrule height #1 \hbox{\vrule width #1
\vbox{#2}\vrule width #1 }\hrule height #1 }}
\def\fillbox#1{\hbox to #1{\vbox to #1{\vfil}\hfil}}
\def\ybox{{\lower 1.3pt \yboxit{0.4pt}{\fillbox{8pt}}\hskip-0.2pt}}

\def\comments#1{}

\def\II{\relax{I\kern-.10em I}}

\def\IZ{\relax\ifmmode\mathchoice
{\hbox{\cmss Z\kern-.4em Z}}{\hbox{\cmss Z\kern-.4em Z}}
{\lower.9pt\hbox{\cmsss Z\kern-.4em Z}}
{\lower1.2pt\hbox{\cmsss Z\kern-.4em Z}}\else{\cmss Z\kern-.4em
Z}\fi}
\def\IB{\relax{\rm I\kern-.18em B}}
\def\IC{{\relax\hbox{$\inbar\kern-.3em{\rm C}$}}}
\def\ID{\relax{\rm I\kern-.18em D}}
\def\IE{\relax{\rm I\kern-.18em E}}
\def\IF{\relax{\rm I\kern-.18em F}}
\def\IG{\relax\hbox{$\inbar\kern-.3em{\rm G}$}}
\def\IGa{\relax\hbox{${\rm I}\kern-.18em\Gamma$}}
\def\IH{\relax{\rm I\kern-.18em H}}
\def\II{\relax{\rm I\kern-.18em I}}
\def\IK{\relax{\rm I\kern-.18em K}}
\def\IP{\relax{\rm I\kern-.18em P}}

%

\def\inbar{\,\vrule height1.5ex width.4pt depth0pt}

\font\cmss=cmss10 \font\cmsss=cmss10 at 7pt
\def\IR{\relax{\rm I\kern-.18em R}}

\ifamsf
\def\IC{\Bbb{C}}
\def\IP{\Bbb{P}}
\def\IR{\Bbb{R}}
\def\IZ{\Bbb{Z}}
\fi

\def\BZ{\IZ}

\def\BC{\IC}

\def\lp10{l_P^{10}}
\def\lp11{l_P^{11}}
\def\R11{R_{11}}

\Title{\vbox{\baselineskip12pt\hbox{hep-th/9711124}\hbox{
CU-TP-869}}}
{\vbox{
\centerline{D-Brane Topology Changing Transitions} }}
\centerline{Brian R. Greene
}
\medskip
\centerline{Departments of Physics and Mathematics}
\centerline{Columbia University }
\centerline{New York, NY 10025}
\centerline{\tt greene@lnsth1.lns.cornell.edu}
\medskip
\bigskip
\noindent
We study D-branes on three-dimensional orbifold backgrounds
that admit topologically distinct resolutions differing
by flop transitions. We show that these distinct phases
are part of the vacuum moduli space of the super Yang-Mills
gauge theory describing the D-brane dynamics. In this way
we establish that D-branes --- like fundamental strings ---
allow for physically smooth topology changing transitions.

\Date{November 1997}
\lref\bss{T. Banks, N. Seiberg, and E. Silverstein, ``Zero and
One-dimensional Probes with $N{=}8$ Supersymmetry,'' hep-th/9703052.}
\lref\DHVW{L. Dixon, J. A. Harvey, C. Vafa, and E. Witten, ``Strings on
Orbifolds, I, II'' Nucl. Phys. B261 (1985) 678; Nucl. Phys. B274 (1986) 285.}
\lref\itoreid{Y. Ito and M. Reid, ``The McKay Correspondence for Finite
Subgroups of SL(3,\BC),'' in: {\it Higher Dimensional Complex Varieties}\/
(M.~Andreatta et al., eds.), de Gruyter, 1996, p.~221; alg-geom/9411010.}
\lref\reid{M. Reid, ``McKay Correspondence,'' alg-geom/9702016.}
\lref\GanMorSei{O. J. Ganor, D. R. Morrison, and N. Seiberg, ``Branes,
Calabi--Yau Spaces, and Toroidal Compactification of the $N{=}1$
Six-Dimensional $E_8$ Theory,'' Nucl. Phys. B487 (1997) 93;
hep-th/9610251.}
\lref\MPorb{D. R. Morrison and M. R. Plesser, to appear.}
\lref\sag{A. Sagnotti, ``Some Properties of Open-String Theories,''
hep-th/9509080.}
\lref\kron{P. B. Kronheimer, ``The Construction of ALE Spaces as
Hyper-K\"{a}hler Quotients,'' J. Diff. Geom.  29 (1989) 665.}
\lref\infirri{A. V. Sardo Infirri, ``Partial Resolutions of Orbifold
Singularities via Moduli Spaces of HYM-type Bundles,'' alg-geom/9610004.}
\lref\infirritwo{A. V. Sardo Infirri, ``Resolutions of Orbifold Singularities
and Flows on the McKay Quiver,'' alg-geom/9610005.}
\lref\polcai{J.~Polchinski and Y.~Cai, ``Consistency of Open Superstring
Theories,'' Nucl. Phys.  B296 (1988) 91.}
\lref\bwb{M. R. Douglas, ``Branes within Branes,'' hep-th/9512077.}
\lref\dm{M. R. Douglas and G. Moore, ``D-Branes, Quivers, and ALE Instantons,''
hep-th/9603167.}
\lref\dg{M. R. Douglas and B. R. Greene, ``Metrics on D-brane Orbifolds'',
hep-th/9707214. }
\lref\JM{C. Johnson and R. Myers, ``Aspects of Type IIB Theory on ALE
Spaces,'' hep-th/9610140.}
\lref\egs{M. R. Douglas, ``Enhanced Gauge Symmetry in M(atrix) Theory,''
hep-th/9612126.}
\lref\polpro{J.~Polchinski, ``Tensors from K3 Orientifolds,''
hep-th/9606165.}
\lref\BFSS{T. Banks, W. Fischler, S. H. Shenker and L. Susskind,
``M Theory as a Matrix Model: A Conjecture,'' hep-th/9610043.}
\lref\ooy{H. Ooguri, Y. Oz and Z. Yin, ``D-Branes on Calabi--Yau Spaces and
Their Mirrors,'' Nucl.Phys. B477 (1996) 407; hep-th/9606112.}
\lref\agm{P. S. Aspinwall, B. R. Greene and D. R. Morrison, ``Calabi--Yau
Moduli Space, Mirror Manifolds and Spacetime Topology Change in
String Theory,'' Nucl. Phys. B416 (1994) 414; hep-th/9309097.}
\lref\rAGMsd{P. S. Aspinwall, B. R. Greene and D. R. Morrison, ``Measuring
Small Distances in $N{=}2$ Sigma Models,''
Nucl. Phys. B420 (1994) 184; hep-th/9311042.}
\lref\aspinwall{P. S. Aspinwall, ``Enhanced Gauge Symmetries and K3
Surfaces,'' Phys. Lett. B357 (1995) 329; hep-th/9507012.}
\lref\dos{M. R. Douglas, H. Ooguri and S. H. Shenker, ``Issues in M(atrix)
Theory Compactification,'' hep-th/9702203.}
\lref\rWP{E. Witten, ``Phases of $N{=}2$ Theories In Two Dimensions,''
Nucl. Phys. B403 (1993) 159; hep-th/9301042.}
\lref\witPT{E. Witten, ``Phase Transitions In M-Theory And F-Theory,''
Nucl. Phys. B471 (1996) 195; hep-th/9603150.}
\lref\fulton{W. Fulton, {\it Introduction to Toric Varieties,}
Princeton University Press, 1993.}
\lref\oda{T. Oda, {\it Convex Bodies and Algebraic Geometry,}
Springer-Verlag, 1988.}
\lref\rGK{B. R. Greene and Y. Kanter, ``Small Volumes in Compactified
String Theory,'' hep-th/9612181.}
\lref\AG{P. S. Aspinwall and B. R. Greene, ``On the Geometric
Interpretation of $N{=}2$ Superconformal Theories,''
Nucl. Phys. B437 (1995) 205; hep-th/9409110.}
\lref\mp{D. R. Morrison and M. R. Plesser, ``Summing the Instantons:
Quantum Cohomology and Mirror Symmetry in Toric
Varieties,'' Nucl. Phys. B440 (1995) 279; hep-th/9412236.}
\lref\rDelzant{T.~Delzant, ``{H}amiltoniens p\'eriodiques et images convexe de
  l'application moment,'' Bull. Soc. Math. France {\bf 116} (1988) 315.}
\lref\rAudin{M.~Audin, {\it The Topology of Torus Actions on Symplectic
Manifolds}, Birkh\"auser, 1991.}
\lref\rCox{D. A. Cox, ``The Homogeneous Coordinate Ring of a Toric
Variety,'' J. Algebraic Geom. 4 (1995) 17; alg-geom/9210008.}
\lref\gp{E. G. Gimon and J. Polchinski,
``Consistency Conditions for Orientifolds and D-Manifolds,'' Phys. Rev. D54
(1996) 1667; hep-th/9601038.}
\lref\GMS{ B. R. Greene, D. R. Morrison, A. Strominger,
``Black Hole Condensation and the Unification of String Vacua'',
  Nucl. Phys. B451 (1995) 109.}
\lref\DKPS{M. R. Douglas, D. Kabat, P. Pouliot, S. H. Shenker,
``D-branes and Short Distances in String Theory'',
Nucl.Phys. B485 (1997) 85-127.}
\lref\DGM{M. R. Douglas, B. R. Greene, D. R. Morrison,
``Orbifold Resolution by D-Branes'', hep-th/9704151.}
\lref\Strom{A. Strominger, ``Massless Black Holes and Conifolds in String Theory'',
Nucl. Phys. B451 (1995) 96.}
\lref\GTASI{B.R. Greene, ``String Theory on Calabi-Yau Manifolds'',
{\it Proceedings of TASI-96},  hep-th/9702155.}
\lref\GString{B.R. Greene, {\it Talk at Strings '97}, Amsterdam, 1997.}
\lref\Polch{ J. Polchinski, ``Dirichlet-Branes and Ramond-Ramond Charges'',
Phys. Rev. Lett. 75 (1995) 4724-4727.}
\lref\DKO{M. R. Douglas, A. Kato, H. Ooguri, ``D-brane Actions on Kahler Manifolds'',
hep-th/9708012.}
\lref\HOV{K. Hori, H. Ooguri, C. Vafa,
``Non-Abelian Conifold Transitions and N=4 Dualities in Three Dimensions'',
hep-th/9705220.}
\lref\GJNS{E. Gava, T. Jayaraman, K. S. Narain, M. H. Sarmadi,
``D-branes and the Conifold Singularity'', Phys.Lett. B388 (1996) 29.}
\lref\JP{D. P. Jatkar, B. Peeters,
``String Theory near a Conifold Singularity'', Phys.Lett. B362 (1995) 73.}
\lref\Muto{T. Muto, ``D-branes on Orbifolds and Topology Change'',
hep-th/9711090.}
%

\newsec{Introduction}

Spacetime --- and the physically accessible
transformations which it can undergo --- is
one arena in which string theory has dramatically changed
our previous conceptions. The discovery of orbifolds 
\DHVW\ showed us that strings can consistently propagate
on mildly singular spaces; $T$-dualities have shown us
that strings can probe identical physics while propagating
through  distinct
backgrounds, and the results of \refs{\agm,\rWP,\GMS}
have shown us that strings allow the topology
of space to change in a physically smooth manner.
Beyond the major shift in our understanding of spacetime
which these results entail, they also provide us
with a clear window on inherently string-based physics
as they are largely
insensitive to many detailed aspects of the theory presently
beyond our analytic control.

More recently, $D$-branes \Polch\ have provided us with a new physical
probe of  short-distance physics in string theory \DKPS.
From this perspective, spacetime is a {\it derived} or
{\it secondary} concept --- emerging from the vacuum moduli
space of $D$-brane world-volume gauge theories. 
Determining how the novel string-based properties of
spacetime mentioned above appear to $D$-branes is
an important and compelling line of study. 
$D$-branes on orbifold backgrounds
have been studied in \refs{\dm,  \JM, \polpro, \DGM, \dg, \dos, \DKO}
and, as we shall discuss in more detail below,
the results match well with
string-derived conclusions. $T$-dualities have been studied in
works that are too numerous to mention
 and many important physical implications have been derived.
On the contrary, even though the conifold
transitions of \GMS\ rely crucially on properties of
wrapped $D$-branes \Strom, and $D$-branes
near conifold singularities have been
subsequently studied in works such as \refs{\HOV, \GJNS, \JP}
 not as much work has been devoted to using
$D$-branes to probe spacetime as it goes through a topology
changing transition. In the present work we focus on this issue. 

In particular, we study the way in which flop transitions ---
so-called mild topology changing processes (see \refs{\agm, \GTASI})
--- appear from
the viewpoint of  $D$-particles. We do this by studying the
$D$-particle vacuum moduli space in an orbifold background
that admits multiple topologically distinct resolutions.
This is  the same philosophy we took in \agm, the difference
being that we now study these rich backgrounds from the perspective
of $D$-branes as opposed to fundamental strings.
We show that these multiple resolutions are directly reflected
in the $D$-brane Yang-Mills vacuum moduli space, and that we
can smoothly interpolate from one topology to another by varying
the Fayet-Illiopoulos parameters that appear in the 
$D$-brane gauge theory.

Specifically, in section 2  we review the results 
of \DGM\
which provide a method for studying $D$-branes on
$\IC^3/\Gamma$, where $\Gamma$ is some finite abelian group. In
section 3,
we recall the novel way in which $D$-branes --- at least
in simple examples --- barely avoid the non-geometrical
phases \refs{\agm,\rWP}
probed by weakly coupled strings, as predicted by the arguments
given in \witPT.
In section 4,
we reiterate the point made in \DGM\ 
that for isolated singularities in
$\IC^3$, the smallest choice for $\Gamma$ giving rise
to flop transitions is the calculationally burdensome case of $\BZ_{11}$.
But we also note the well known fact, as
we stressed in \GString,
that if one foregoes the 
requirement of isolated singularities, then the more manageable case
of $\Gamma = \BZ_2 \times \BZ_2$ is sufficiently rich to admit
distinct resolutions differing by flops.
We then carry out the analysis in this example, and explicitly
show that the $D$-brane vacuum moduli space  incorporates
these flop transitions, thereby showing that this novel piece
of string physics has a direct $D$-brane counterpart.
In section 5 we offer some conclusions.

After this work was completed, a paper by Muto
\Muto\
appeared which took up the challenge of \DGM\ and claims to
have successfully worked out the $\BZ_{11}$ case.

\newsec{D-branes on $\IC^3/\Gamma$}

As in \DGM\
we study a single $D$-brane in type II string theory rolling
around on $\IC^3/\Gamma$.
In the $d = 4$ `non-compact' dimensions,
this has ${\cal N} = 2$ supersymmetry in the closed string sector, 
while in the open string sector it has ${\cal N} = 1$ supersymmetry.
We realize the latter by working with an
${\cal N} = 4$ $d = 4$ theory of $|\Gamma|$ $D$-branes on
the $\IC^3$ covering space,
and then projecting to $\IC^3/\Gamma$. This projection is
defined by making two choices: the action of $\Gamma$ on $\IC^3$ and
the action of $\Gamma$ on the Chan-Paton factors of the
$U(|\Gamma|)$ gauge group. In order to project to a single
$D$-brane in the quotient theory, we should take the latter
action of $\Gamma$ to be the regular representation. The only
requirement we impose on the former action is that it lie
in $SU(3)$ so that at least a single supersymmetry is preserved
in passing to the quotient space.

To find the fields in the quotient theory, we identify
those which are invariant under these combined actions. Explicitly,
if we let $g \in \Gamma$, and denote its actions
on $\IC^3$ and on the Chan-Paton factors by $R(g)$ and $S(g)$
respectively, we can determine the surviving fields as follows:
The components of 
gauge fields $A$, represented by $U(|\Gamma|)$ matrices
with indices along the brane world volume, survive
if
\eqn\gaugeproj{
A_{ij} = ( S(g) A S(g)^{-1})_{ij}.}
The scalar fields living on the brane world-volume --- again
originating as $U(|\Gamma|)$ matrices $X$ --- survive if
\eqn\matterpro{
X^{\alpha}_{ij} = (R(g) S(g) X^{\alpha} S(g)^{-1})_{ij}}
where $R(g)$ only acts non-trivially in the $\IC^3$ directions.

After solving these projection constraints, we  substitute
the surviving fields into the dimensional reduction onto
the brane world volume of the covering space ${\cal N} = 4$
super-Yang-Mills theory, augmented by the Fayet-Illiopoulos
parameters discussed in \dm.
The vacuum moduli space
of this theory --- which we identify with the internal
part of spacetime as probed by the $D$-brane --- arises from
solving the $F$ and $D$-flatness conditions of this supersymmetric
gauge theory. The $F$-flatness constraints come from the
superpotential
$W = Tr [X^1, X^2]X^3$ whose minimization
 $\partial W / \partial X_{ij}^{\alpha}$ yields the equations
\eqn\e{
[X^{\alpha}, X^{\beta}] = 0.}
The $D$-flatness conditions arise from the unbroken
$U(1)$ gauge symmetries that survive the separating of
the $D$-branes and are of the usual form
\eqn\Dterms{
\sum q_{(k) \alpha}^{\beta} |X^{\alpha}_{\beta}|^2 - \zeta_k = 0}
where  $q_{(k) \alpha}^{\beta}$ is the charge of $X^{\alpha}_{\beta}$
under the $k^{\rm th}$ $U(1)$ and $\beta$ runs over the matrix
indices of all surviving components of $X^{\alpha}$.
For ease of notation, we shall temporarily call these surviving
components $x_1,x_2,...,x_{3 |\Gamma|}$ and we shall call the
$|\Gamma| - 1$ by $|\Gamma| + 2$ matrix containing these charges $V$.

In \DGM,
a convenient method for solving these vacuum configuration constraints
was introduced, making use of toric geometry to phrase them all
in a single unified framework. Specifically, the $F$-flatness
conditions are explicitly solved for  $2 |\Gamma| - 2$
of the $x_i$ in terms of the remaining $|\Gamma| + 2$ variables.
These relations can be summarized by writing
\eqn\e{
x_i = \prod_{j = 1}^{|\Gamma| + 2} x_j^{a_{ij}} }
where we assume that the $x$'s are arranged so that the
first $|\Gamma| + 2$ are the independent variables. We then
arrange this data in a $3 |\Gamma|$ by $|\Gamma| + 2$ matrix
$A$ whose entries are the $a_{ij}$. This matrix encodes the
`M-lattice' data (in the language of \agm\ for example)
of the toric variety ${\cal T}$ of the scalar fields that solve
the superpotential constraints.

This takes us part way toward the vacuum moduli space ${\cal V}$ ---
what remains is to impose the $D$-term constraints. Now,
as in \Dterms, $D$-term constraints naturally take the
form of symplectic quotients: we impose a $\IC^*$ constraint
by enforcing  moment map conditions and then modding-out by
a $U(1)$ symmetries. In order to impose these conditions in a
manner that efficiently meshes with our $F$-flatness constraints,
we are therefore led to rephrase the latter in a symplectic
framework as well. This is straightforward to do. From the matrix
$A$ we construct a `dual' $|\Gamma| + 2$ by $k$ matrix  $T$ 
whose columns span the
dual cone to that spanned by the rows of $A$.
(We are not aware of a general expression for $k$; in examples,
it is calculated by brute force enumeration of the dual cone.)
The $k$ columns of $T$ are associated with $k$ homogeneous
coordinates $p_0, p_1,...,p_{k-1}$ as discussed, for example,
in \agm.
 The transpose
of the kernel of $T$, as shown in \AG\
gives the charge matrix $Q$ for ${\cal T}$ realized as
a symplectic quotient. All that remains, therefore, is to
augment $Q$ with the additional symplectic quotients that
enforce the $D$-term constraints.  In \DGM\
it was shown that this is accomplished as follows.

We introduce a $|\Gamma| + 2$ by $k$ matrix $U$ that satisfies
$ TU^t = {\bf 1}$. Then, the $D$-term charges of the fields in this dual
representation are given by the $|\Gamma| - 1$ by $k$ matrix $VU$.
Thus, we can combine the $F$ and $D$-flatness conditions by
concatenating the matrix $Q$ and the matrix $VU$ into a $(k-3)$ by
$k$ matrix
$Q^{\rm total}$. The vacuum moduli space ${\cal V}$ is thus
realized by the symplectic quotient of $\IC^{k}$ by $U(1)^{k-3}$,
with the latter action being determined by $Q^{\rm total}$.

This is almost, but not quite, the whole story. The symplectic
quotient just alluded to involves $k - 3$ moment maps whose precise
form also relies on the values of Fayet-Illiopoulos parameters
coming from the $D$-terms. The choice of these parameters
fills out the definition of the model and allows complete
specification of ${\cal V}$.

\newsec{An Example}

An example will make this  discussion more clear as
well as provide a jumping off point for our discussion of
flops in the next section. We choose the simplest example
of $\Gamma = \BZ_3$, with
action on $\IC^3$ given by
\eqn\e{
R(g): (X,Y,Z) \rightarrow (\omega^{-1} X, \omega^{-1} Y, \omega^{-1} Z)}
where $g$ generates $\BZ_3$,
 $\omega$ is a nontrivial cube-root of unity,
and $S(g)$ is taken to be the regular representation.
Then,
as originally discussed in \DGM, if we call the $9$ variables
$x_1,...,x_9$ = $x_0,x_1,x_2,y_0,y_1,y_2,z_0,z_1,z_2$ (where
the notation $x_{0,1,2}$ denotes the three components of
the coordinate $X$ --- after it is promoted to a
Yang-Mills matrix --- which survive the projection, etc.)
we have the matrix $A$ given by the following integer entries

\eqn\coneM{\matrix{
&&x_0&y_0&z_0&z_1&z_2 \cr
x_0&&1&0&0&0&0 \cr
x_1&&1&0&-1&1&0 \cr
x_2&&1&0&-1&0&1 \cr
y_0&&0&1&0&0&0 \cr
y_1&&0&1&-1&1&0 \cr
y_2&&0&1&-1&0&1 \cr
z_0&&0&0&1&0&0 \cr
z_1&&0&0&0&1&0 \cr
z_2&&0&0&0&0&1, }
}
where we label the columns by the chosen set of independent fields.
Its dual $T$ is
\eqn\coneN{
T=\pmatrix{
1 & 0 & 0 & 1 & 0 & 0 \cr
0 & 1 & 0 & 1 & 0 & 0 \cr
0 & 0 & 1 & 1 & 0 & 0 \cr
0 & 0 & 1 & 0 & 1 & 0 \cr
0 & 0 & 1 & 0 & 0 & 1}
}
which has kernel given by the transpose of
\eqn\echargmat{
Q=\pmatrix{
1 & 1 & 1 & -1 & -1 & -1
}.}
A suitable choice for the matrix $U$ is
\eqn\eU{
U=\pmatrix{
0 & -1 & -1 & 1 & 1 & 1 \cr
0 & 1 & 0 & 0 & 0 & 0 \cr
0 & 0 & 1 & 0 & -1 & -1 \cr
0 & 0 & 0 & 0 & 1 & 0 \cr
0 & 0 & 0 & 0 & 0 & 1
}}
which yields the matrix
\eqn\efull{
Q^{\rm total}=\pmatrix{
1 & 1 & 1 & -1 & -1 & -1 \cr
0 & 0 & 0 & 0 & -1 & 1 \cr
0 & 0 & 0 & 1 & 0 & -1
}.}

As in the last section, from $Q^{\rm total}$  and a specification
of the Fayet-Illiopoulos parameters we can determine the precise
form of ${\cal V}$. However, even without specifying values for
the Fayet-Illiopoulos parameters (which we will do momentarily)
we can use $Q^{\rm total}$ to determine the `N-lattice' data
of ${\cal V}$ by taking the  kernel of $Q^{\rm total}$ and
eliminating all redundant rows. Doing so gives us the point set
in $\IR^3$:
\eqn\pointszthree{\eqalign{
 e_1 &=( 1, 0, 0) \cr
 e_2 &=(0, 1, 0) \cr
 e_3 &=( 0, 0, 1) \cr
 e_4 &=(-1, -1, 3)},
} 
which we immediately recognize as the toric
data for resolving $\IC^3/ \BZ_3$. Thus, strikingly, the
$D$-brane vacuum moduli space ${\cal V}$ aligns with
the internal space on which we set the $D$-brane moving.

A more subtle test of this correspondence 
 requires that we also
include the Fayet-Illiopoulos terms, as we now do.
As in \DGM,
a convenient way to do this is to augment $Q^{\rm total}$
by an additional column which encodes the Fayet-Illiopoulos
term for each of the corresponding rows. For this example
this yields

\eqn\erewrite{
\pmatrix{
1 & 1 & 1 & -1 & -1 & -1 & 0 \cr
0 & 0 & 0 & 0 & -1 & 1 & {{ \zeta}_{1}} \cr
0 & 0 & 0 & 1 & 0 & -1 & {{ \zeta}_{2}}
}.}

Notice that the Fayet-Illiopoulos term vanishes for the upper
row since this constraint originated from an $F$-flatness
condition in our original model. This, in fact, is the crucial
point. It means that the symplectic quotient associated with
$Q^{\rm total}$ is {\it not} generic since the moment maps
do not have generic arguments. This, in turn, constrains the
regions of the phases-picture of \refs{\agm, \rWP}
that are physically accessible by the D-brane. Lets recall 
what this observation implies in this simple example.

Doing so requires that we allow $\zeta_{1,2}$ to freely run
over all (real) values, and for each such possibility we must carry
out the symplectic quotient to determine ${\cal V}$.
This is easier done than said: a choice of sign for
each of $\zeta_{1,2}$, via the last two rows of 
$Q^{\rm total}$, ensures that two of the last three
$p$ variables {\it cannot} vanish. We then eliminate
these two variables by doing row operations that
eliminate them from the first row. For example,
assume that $\zeta_{1,2}$ are both positive. Then we
see that $p_3$ and $p_5$ cannot vanish and we eliminate
them by adding row three and two times row two to row one.
This yields the matrix
\eqn\esolve{
\pmatrix{
1 & 1 & 1 & 0 & -3 & 0 & 2\,{{ \zeta}_{1}} + {{ \zeta}_{2}} \cr
0 & 0 & 0 & 0 & -1 & 1 & {{ \zeta}_{1}} \cr
0 & 0 & 0 & 1 & -1 & 0 & {{ \zeta}_{1}} + {{ \zeta}_{2}}
}.}

The important thing to note is that the Fayet-Illiopoulos term
in the first row, $2 \zeta_1 + \zeta_2$, is of {\it fixed}
(positive) sign. This ensures that we are in the blown up
phase of $\IC^3/\BZ_3$. A similar thing holds true for
all other choices of signs of $\zeta_{1,2}$, as the
reader can easily confirm. In this simple
example, then, carrying out all but $h^{11} = 1$ of the symplectic
quotients --- the ones which manifestly constrain some
of the homogeneous coordinates to lie in $\IC^*$ ---
uniquely fixes the sign of the remaining Fayet-Illiopoulos
parameters. And moreover, the constraint keeps us in the fully
resolved geometric phase of the model, not allowing us
to enter the analog of the Landau-Ginsburg phase.
This is in precise keeping with the observation of
\witPT\ that the results of \rAGMsd\ can be used to
argue that only geometric phases survive to the
strong coupling limit. As $D$-branes are the most
relevant degrees of freedom in this limit, we anticipate
that they should only probe geometric phases as well,
and this is precisely what occurs in this example.

The question we now pose is whether $D$-branes necessarily
probe {\it all} possible geometric phases of a model.
This example, and the others studied in \DGM\ are
too simple to address this issue, as they each have
a unique geometric phase region, and hence we
must undertake further analysis.

\newsec{D-brane Flops}

As mentioned at the end of \DGM,
the simplest cyclic quotient singularity of $\IC^3$ that
admits flops is $\BZ_{11}$, a prohibitively large number.
So rather than examining that case, let's leave the domain
of isolated cyclic quotients and consider the case of $\Gamma = \BZ_2 \times
\BZ_2$. As is well known, $\IC^3/(\BZ_2 \times \BZ_2)$ does admit
distinct flop related resolutions. And as mentioned in
\GString\
and now explicitly shown, unlike the case reviewed above,
there is {\it freedom in the sign} of the Fayet-Illiopoulos
parameters that arise from reducing $Q^{\rm total}$ by eliminating
homogeneous variables forced to lie in $\IC^*$. As we shall see,
this freedom
coincides with the ability to flop rational curves in
the $D$-brane vacuum moduli space ${\cal V}$.

One complication of this quotient is that the resulting singularity
is not isolated. For the analysis we perform here, this 
hardly changes our procedure. However, if we wanted to fully
express the complexified conformal field theory blow up modes in terms of
the Yang-Mills gauge theory parameters, we would have to work
harder than we do here. This is an issue to which
we hope to  return. 

We choose our $\BZ_2 \times \BZ_2$ generators $g_1, g_2$ to act
on $\IC^3$ according to
\eqn\egenerators{
g_1: (X,Y,Z) \rightarrow (-X, -Y, Z)}
\eqn\egeneratorss{
g_2: (X,Y,Z) \rightarrow (-X, Y, -Z)}
Then, after diagonalizing, the regular representation is given
by
\eqn\eregone{
S(g_1)=\pmatrix{
1 & & & \cr
  &1& & \cr
  & &-1& \cr
  & & &-1 }
}
and
\eqn\eregtwo{
S(g_2)=\pmatrix{
1 & & & \cr
  &-1& & \cr
  & &1& \cr
  & & &-1 }.
}

It is now straightforward to do the projections and realize
that the surviving fields are: 

\eqn\efx{
(x_1,x_2,x_3,x_4) = (X_{14}, X_{23}, X_{32}, X_{41}) }
\eqn\efy{
(y_1,y_2,y_3,y_4) = (Y_{13}, Y_{24}, Y_{31}, Y_{42}) }
\eqn\efz{
(z_1,z_2,z_3,z_4) = (Z_{12}, Z_{21}, Z_{34}, Z_{43}) }
where the right-hand-side denotes, again, the components of
the matrices arising from the Yang-Mills description (whose
spacetime index is in the direction of the coordinates $X,Y$ or $Z$).

A little algebra now shows that the $F$-flatness constraints
are solved by the relations encoded in the matrix $A$:

\eqn\coneMTwotwo{\matrix{
&&x_1&x_2&x_3&y_1&y_2&z_1 \cr
x_1&&1&0&0&0&0&0 \cr
x_2&&0&1&0&0&0&0 \cr
x_3&&0&0&1&0&0&0 \cr
x_4&&-1&1&1&0&0&0 \cr
y_1&&0&0&0&1&0&0 \cr
y_2&&0&0&0&0&1&0 \cr
y_3&&-1&0&1&0&1&0 \cr
y_4&&-1&0&1&1&0&0 \cr
z_1&&0&0&0&0&0&1 \cr
z_2&&-1&1&0&-1&1&1 \cr
z_3&&0&0&0&-1&1&1 \cr
z_4&&-1&1&0&0&0&1 }
}

Following the procedure discussed in section 2,  we now
calculate the dual matrix $T$:
\eqn\coneNtwotwo{
T=\pmatrix{
1 & 1 & 1 & 0 & 0 & 0 & 0 & 0 & 0 \cr
0 & 1 & 1 & 1 & 0 & 0 & 0 & 0 & 0 \cr
1 & 1 & 0 & 0 & 1 & 0 & 0 & 0 & 0 \cr
0 & 0 & 1 & 0 & 0 & 1 & 0 & 1 & 0 \cr
0 & 0 & 1 & 0 & 0 & 1 & 0 & 0 & 1 \cr
1 & 0 & 0 & 0 & 0 & 0 & 1 & 1 & 0}.
}
The kernel of this matrix is
\eqn\ekerneltwo{
Q=\pmatrix{
0 & 1 &-1 & 0 &-1 & 1 & 0 & 0 & 0 \cr
-1& 1 & 0 &-1 & 0 & 0 & 1 & 0 & 0 \cr
-1& 2 &-1 &-1 &-1 & 0 & 0 & 1 & 1}
.}
This gives us our standard toric geometry (equivalently, linear sigma model)
construction of what we have been calling ${\cal T}$ --- the
locus of the fields meeting the superpotential constraints.
We now need to include the $D$-term constraints. To do so, we
follow the procedure given earlier and calculate the matrix
$U$ to be
\eqn\Umatrix{
U=\pmatrix{
0 & 1 & 0 &-1 &-1 & 0 & 0 & 0 & 0 \cr
0 & 0 & 0 & 1 & 0 & 0 & 0 & 0 & 0 \cr
0 & 0 & 0 & 0 & 1 & 0 & 0 & 0 & 0 \cr
-1& 1 & 0 &-1 & 0 & 0 & 0 & 1 & 0 \cr
1 &-2 & 1 & 1 & 1 & 0 & 0 &-1 & 0 \cr
1 &-1 & 0 & 1 & 0 & 0 & 0 & 0 & 0}.
}

Together with the $U(1)^3$ charge matrix
\eqn\Vmatrix{
V=\pmatrix{
1 & 0 & 0 & 1 & 0 & 1 \cr
0 & 1 &-1 & 0 & 1 &-1 \cr
0 &-1 & 1 &-1 & 0 & 0},
}
we learn that our 9 homogeneous variables,
$p_0, p_1,...,p_8$ have charges $VU$
given by
\eqn\Vmatrix{
VU=\pmatrix{
0 & 1 & 0 &-1 &-1 & 0 & 0 & 1 & 0 \cr
0 &-1 & 1 & 1 & 0 & 0 & 0 &-1 & 0 \cr
1 &-1 & 0 & 0 & 1 & 0 & 0 &-1 & 0}.
}
Concatenating this with the charge matrix $Q$ yields
\eqn\Qtotmatrix{
Q^{\rm total}=\pmatrix{
0 & 1 &-1 & 0 &-1 & 1 & 0 & 0 & 0 \cr
-1& 1 & 0 &-1 & 0 & 0 & 1 & 0 & 0 \cr
-1& 2 &-1 &-1 &-1 & 0 & 0 & 1 & 1 \cr
0 & 1 & 0 &-1 &-1 & 0 & 0 & 1 & 0 \cr
0 &-1 & 1 & 1 & 0 & 0 & 0 &-1 & 0 \cr
1 &-1 & 0 & 0 & 1 & 0 & 0 &-1 & 0}.
}

Now, before including the Fayet-Illiopoulos terms, we calculate
the kernel of this charge matrix, and find the (distinct) points
in $\IR^3$ 
\eqn\ePoints{\eqalign{
e_1& =  (1, 0, 0)\cr
e_2& =  (0, 1, 0)\cr
e_3& =  (0, 0, 1)\cr
e_4& =  (2, 1, -2)\cr
e_5& =  (0, -1, 2)\cr
e_6& =  (1, 1, -1)\cr}}

It is not hard to see that this {\it is} the toric point set
for resolving $\IC^3/(\BZ_2 \times \BZ_2)$. Explicitly,
the toric data for $\IC^3/(\BZ_2 \times \BZ_2)$ can be found by
augmenting the $\BZ^{\oplus 3}$ lattice (generated by the
standard basis vectors $e_1,e_2,e_3$ over $\BZ$) by the fractional
points $(1/2, 1/2, 0)$ and $(1/2, 0, 1/2)$. In other words,
we build a new lattice whose generators are these points,
together with those generating the original $\BZ^{\oplus 3}$ lattice.
The reason we do this,
as explained for instance in \AG,
is that, in general,
including fractional points $(h_1,...,h_n)$ in the toric
$N$ lattice $\BZ^{\oplus n}$ yields the one-parameter group
action on $(\IC^*)^n$ with coordinates $(\rho_1,...,\rho_n)$
of the form

\eqn\emap{
(\rho_1,...,\rho_n) \rightarrow (e^{2\pi i h_1}\rho_1,...,e^{2\pi i h_n}\rho_n).}
For $h_i = a_i/n$ with $a_i \in \BZ$, we see that this enforces
a $\BZ_n$ identification, that is, a $\BZ_n$ orbifold.  Thus, adding the
points $(1/2, 1/2, 0)$ and $(1/2, 0, 1/2)$ enforces the identifications
on $\IC^3$ given by the action of $g_1$ and $g_2$. Note also that
we necessarily must also include $(0,1/2, 1/2)$ as this identification
arises from $g_1 g_2$ (and, equivalently, it
 is also a lattice point in our augmented
lattice). To put this in a more
recognizable form,  we can now change basis to restore integrality in
our augmented lattice. Namely, with

\eqn\ePointsfrac{\eqalign{
e_1& = (1, 0, 0)\cr
e_2& =  (0, 1, 0)\cr
e_3& = (0, 0, 1)\cr
e'_4& =  (1/2, 1/2, 0)\cr
e'_5& =  (1/2, 0, 1/2)\cr
e'_6& =  (0, 1/2, 1/2)\cr}}
we can change basis by expressing all vectors in terms
of $e_2, e'_4, e'_5$, which we now call $(0, 1, 0), (0, 0, 1), (1, 0, 0 )$,
yielding $e_1 = (0, -1, 2), e_3 = (2, 1, -2)$ and $e'_6 = (1, 1, -1)$.
We recognize that this list of six points in $\IR^3$ exactly matches
the list of points in \ePoints, thus establishing
that we have recovered --- from our $D$-brane vacuum moduli space ---
the toric data for resolving our internal $\IC^3/(\BZ_2 \times \BZ_2)$
component of space.

For the final step we need to include the Fayet-Illiopoulos parameters.
Doing so yields
\eqn\QtotmatrixFI{
Q^{\rm total}=\pmatrix{
0 & 1 &-1 & 0 &-1 & 1 & 0 & 0 & 0 & 0 \cr
-1& 1 & 0 &-1 & 0 & 0 & 1 & 0 & 0 & 0 \cr
-1& 2 &-1 &-1 &-1 & 0 & 0 & 1 & 1 & 0 \cr
0 & 1 & 0 &-1 &-1 & 0 & 0 & 1 & 0 & \zeta_1 \cr
0 &-1 & 1 & 1 & 0 & 0 & 0 &-1 & 0 & \zeta_2 \cr
1 &-1 & 0 & 0 & 1 & 0 & 0 &-1 & 0 & \zeta_3}.
}
Then, by adding the fourth row to each of rows five and six this
becomes
\eqn\Qtotrewrite{
\pmatrix{
0 & 1 &-1 & 0 &-1 & 1 & 0 & 0 & 0 & 0 \cr
-1& 1 & 0 &-1 & 0 & 0 & 1 & 0 & 0 & 0 \cr
-1& 2 &-1 &-1 &-1 & 0 & 0 & 1 & 1 & 0 \cr
0 & 1 & 0 &-1 &-1 & 0 & 0 & 1 & 0 & \zeta_1 \cr
0 & 0 & 1 & 0 &-1 & 0 & 0 & 0 & 0 &  \zeta_1 + \zeta_2 \cr
1 & 0 & 0 &-1 & 0 & 0 & 0 & 0 & 0 &  \zeta_1 + \zeta_3},
}
which is a particularly nice form since it immediately allows us
to identify homogeneous coordinates lying in $\IC^*$ by the signs
of $\zeta_1 + \zeta_2$ and $\zeta_1 + \zeta_3$. In fact, for definiteness
let's assume that we examine the part
of moduli space in which each of these combinations of $\zeta$'s
is positive. This means we must eliminate $p_0$ and $p_2$.
We can again accomplish this by row operations taking us to
\eqn\Qtotrewritea{
\pmatrix{
0 & 1 & 0 & 0 &-2 & 1 & 0 & 0 & 0 & \zeta_1 + \zeta_2 \cr
0 & 1 & 0 &-2 & 0 & 0 & 1 & 0 & 0 &  \zeta_1 + \zeta_3 \cr
0 & 0 & 0 & 0 & 0 & 0 & 0 &-1 & 1 &   \zeta_2 + \zeta_3 \cr
0 & 1 & 0 &-1 &-1 & 0 & 0 & 1 & 0 & \zeta_1 \cr
0 & 0 & 1 & 0 &-1 & 0 & 0 & 0 & 0 &  \zeta_1 + \zeta_2 \cr
1 & 0 & 0 &-1 & 0 & 0 & 0 & 0 & 0 &  \zeta_1 + \zeta_3},
}
where we have also taken the liberty of doing invertible row operations
that simplify row 3, as shown.
This is an even nicer form since if we further assume, for instance,
 that
$\zeta_2 + \zeta_3$ is positive, the third row allows
us to eliminate $p_8$ (which does not appear in any other
constraint equation). Putting all this together takes us
to the reduced form of a symplectic quotient of $\IC^6$
(the variables $p_1,p_3,p_4,p_5,p_6,p_7$) by the
$U(1)^3$ action given by

\eqn\Finalform{
Q^{\rm reduced}=\pmatrix{
1 & 0 &-2 & 1 & 0 & 0 & \zeta_1 + \zeta_2 \cr
1 &-2 & 0 & 0 & 1 & 0 & \zeta_2 + \zeta_3 \cr
1 &-1 &-1 & 0 & 0 & 1 & \zeta_1},
}
in which we have now used rows 3, 5, and 6 of \Qtotrewritea,
together with our assumption on the 
$\zeta$'s,
to eliminate the columns associated with
$p_0, p_2$, and $p_8$.

The key point to  notice is that the part of moduli space
we are focusing on, $\zeta_1 + \zeta_2 > 0, \zeta_1 + \zeta_3 > 0,
\zeta_2 + \zeta_3 > 0$, does {\it not} constrain the sign
of $\zeta_1$ in the third row of \Finalform\ above. That is, unlike
the case of $\BZ_3$ reviewed earlier, when we fix a region of moduli
space to eliminate constrained homogeneous variables in this context,
it does not uniquely constrain the sign of the resulting combination
of Fayet-Illiopoulos parameters.

The reason this is of relevance comes from
the toric interpretation of the last line in \Finalform.
This equation is
\eqn\e{
|p_1|^2 + |p_7|^2 - |p_3|^2 - |p_4|^2 = \zeta_1}
which yields a $\IP^1$ that is flopped as the sign of
$\zeta_1$ changes. In toric terms, this equation is
equivalent to the data of four vectors
in $\IR^3$, $\vec u_2,\vec u_7,\vec u_3, \vec u_4$ satisfying
\eqn\e{
\vec u_1 + \vec u_7 = \vec u_3 + \vec u_4,} as shown in figure 1.

\ifig\figone{The four toric vectors $\vec u_i$ in $\IR^3$ yielding
a  (double-point)  singularity.}
{\epsfxsize1.0in
\epsfbox{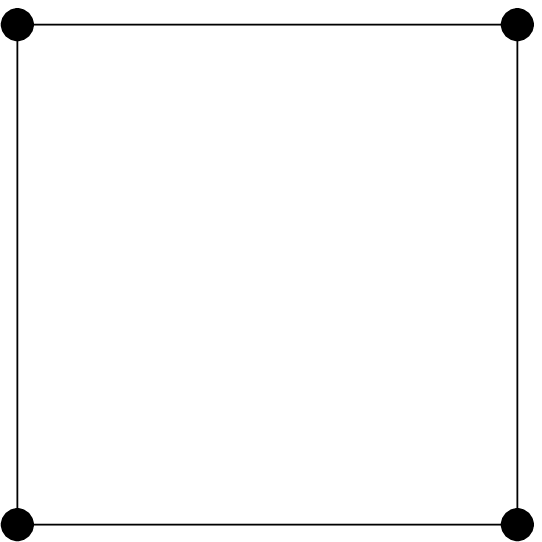}}

As written, these toric points yield an ordinary double point.
We can resolve this singularity by triangulating the point set,
but, as we illustrate in figure 2, this can be done in two ways, 
differing by a flop of a rational curve. From a symplectic
standpoint, it is the sign of $\zeta_1$ that determines which
of these two resolutions is used.

\ifig\figone{The two small resolutions of the singularity.}
{\epsfxsize2.0in
\epsfbox{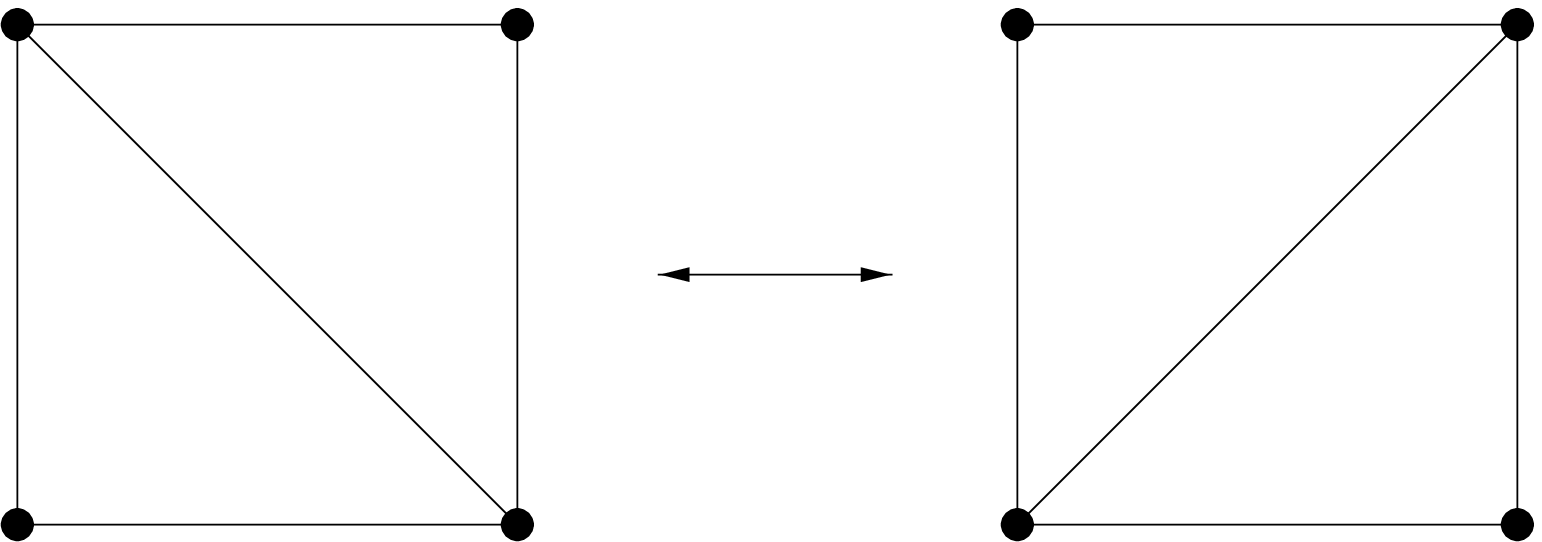}}

And so we see that the freedom of sign in $\zeta_1$ corresponds
to the freedom of flopping a rational curve in the $D$-brane
vacuum moduli space. In the full context of $\IC^3/ (\BZ_2 \times \BZ_2)$
we see these flops in the toric diagram of the points in
\Finalform\ of which we show one example in figure 3.

\ifig\figone{An example of a flop transition in
resolving $\IC^3/ (\BZ_2 \times \BZ_2)$. }
{\epsfxsize3.5in
\epsfbox{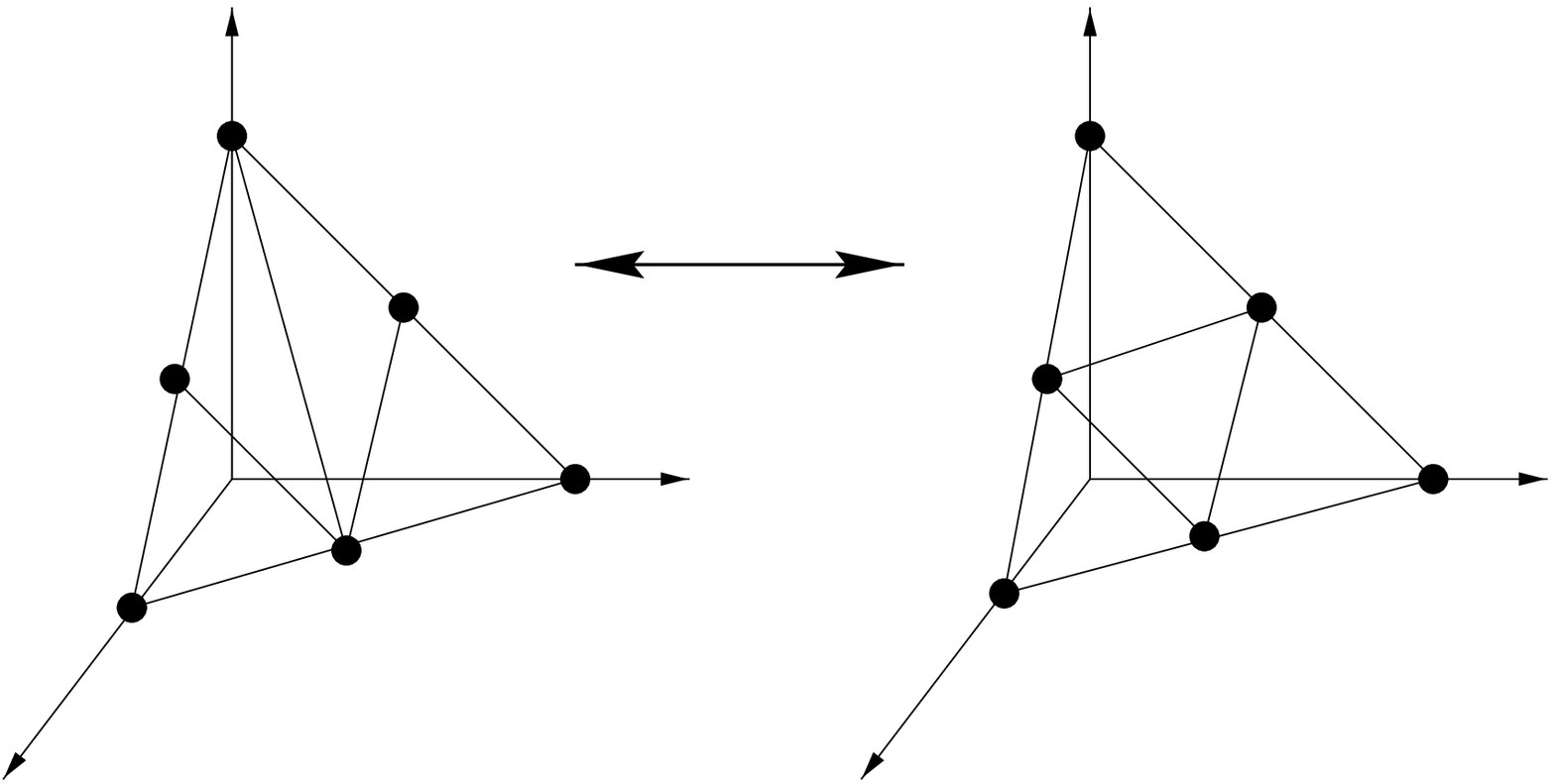}}

\newsec{Conclusions}

The arguments presented here are closer in spirit to those of
\agm\ than to those of \rWP\ because they establish
topology change from a global as opposed to a local perspective.
We have  shown that the $D$-brane vacuum moduli space does have
regions related by flop transition, and then, by virtue of
supersymmetry, we know that singularities are at least of complex
codimension 1, ensuring smooth passage from one region to another.
Nevertheless, it would be very interesting to understand
the  $D$-brane analog of the analysis of
\rWP\ as this would give insight into the microscopic
features underlying topology change according to these
non-perturbative string degrees of freedom.

Furthermore, it would be very interesting to fully understand
the map between the {\it complexified} nonlinear sigma model variables
and the parameters of the $D$-brane gauge theory (which, in the
non-isolated case, apparently
requires additional gauge theory parameters). This would
allow direct comparison of quantum volumes as probed by strings
and by $D$-branes, and, for instance, could be used to study
the local dynamics around the more drastic conifold topology
changing processes as well.

\bigskip
\centerline{{\bf Acknowledgments}}\nobreak

The author gratefully acknowledges
the support of a National Young Investigator Award, the
Department of Energy, and
an Alfred P. Sloan Foundation Fellowship.

\listrefs
\end